\documentclass[aps,prl,twocolumn,superscriptaddress,showpacs]{revtex4-2}

\usepackage{amsmath,amssymb,amsfonts,bm}
\usepackage{graphicx}
\makeatletter\def\Gin@extensions{.pdf,.png,.jpg}\makeatother
\graphicspath{{./}{figures/}}
\usepackage[colorlinks=true,linkcolor=blue,citecolor=blue]{hyperref}
\usepackage{xcolor}
\newcommand{\bbe}{\begin{equation}}
\newcommand{\bbq}{\end{equation}}
\newcommand{\bea}{\begin{eqnarray}}
\newcommand{\eea}{\end{eqnarray}}

\newcommand{\ads}{\text{AdS}}

\newcommand{\Dsig}{\Delta_\sigma}
\newcommand{\Deps}{\Delta_\epsilon}
\newcommand{\vsig}{\mathcal{V}_\sigma}
\newcommand{\veps}{\mathcal{V}_\epsilon}

\begin{document}

\title{Learning  the Emergent Bulk Geometry of the Three Dimensional Ising Model}
\author{Ritam Basu}
\email{ritam.basu@tifr.res.in}
\affiliation{Department of Theoretical Physics, Tata Institute of Fundamental
Research, 1 Homi Bhabha Road, Mumbai 400005, India}
\date{\today}

\begin{abstract}
%% ---------------------------------------------------------------
%% Fill the two numbers from the final sweep. Keep it to 5 sentences.
%% ---------------------------------------------------------------
We use Monte Carlo data for the three dimensional Ising model at critical temperature  to predict  an emergent radial geometry of the bulk  using machine learning. We use a differentiable solver for a probe field on an asymptotically $\ads_4$ background. It is trained to reproduce the measured momentum space two point functions of the boundary operator ( $\sigma$ and $\epsilon$ ) . We find a  single metric cannot serve both the probe field operators. Now fitting each channel separately improves its own residual by a factor of three to five but degrades the other by up to two orders of magnitude. We thus  quantify this tension between this two field operators and show that it tension survives changes of the momentum window and of the lattice size. In this way , we numerically show  that the Ising bulk is not captured by a single classical metric for a theory like 3d Ising model with central charge of $O(1)$. 
\end{abstract}
\maketitle

%% ==================================================================
\textit{Introduction.}---%
We know  from the holographic dictionary that a conformal field theory  maps to a gravitational theory in one higher dimension\cite{Maldacena:1997re,
Gubser:1998bc, Witten:1998qj}.
For theories with a large central charge and a sparse spectrum the bulk is expected to be Einstein gravity \cite{Heemskerk:2009pn}. But for a boundary
theory with a central charge of $O(1)$ no such expectation holds. Specially $3d-$Ising model can be thought of the sharpest example of it. Its critical exponents are known to high precision from the conformal bootstrap \cite{Kos:2016ysd, SimmonsDuffin:2016wlq, Chang:2024whx}. But Its gravity dual is not known.

Two lines of thought exist in the literature. 
The large $N$ vector models are dual to higher spin gravity \cite{Klebanov:2002ja, Giombi:2009wh}.
Their higher spin symmetry is exact only for free theories \cite{Maldacena:2011jn} and is weakly broken at large $N$ \cite{Alday:2015ega}.
But the Ising point sits at $N=1$ where the breaking is not weak. Alternatively the model has long been conjectured to be a strongly coupled fermionic string \cite{Polyakov:1987ez, Iqbal:2020msy}. But none of this picture yields a bulk one can solve.

In this project we take a different route.
We do not assume a bulk but approch this problem from the boundary CFT. We measure boundary correlators on the lattice and then ask what geometry, if anything possible , can to reproduce them. Machine learning methods for this inverse problem were introduced in \cite{Hashimoto:2018ftp, Hashimoto:2018bnb} and developed with neural
differential equations in \cite{Hashimoto:2020jug}.
But they have been applied this technique to holographic QCD \cite{Hashimoto:2018bnb},
to transport in strange metals \cite{Li:2022zjc, Ahn:2024gjf, Ahn:2025tvo}, and linear response \cite{Hashimoto:2024ymt}. In all of these cases a single field is fitted. But here we fit two operators at once and use the mismatch as the observable.

%% =====================================================================
\textit{Setup.}---%
We work with static planar ansatz in 4-d bulk, 
\bbe
ds^2 = \frac{L^2}{z^2}\Big( f(z)\, d\tau^2 + \frac{dz^2}{g(z)} + dx^2 + dy^2 \Big),
\label{eq:ansatz}
\bbq
with $f,g \to 1$ as $z \to 0$. We then set $L=1$.
We probe a scalar filed of mass $m$ in the bulk which obeys the Klein Gordon equation on this background. In Fourier space this is a radial ordinary differential equation,
\bbe
\partial_z\!\Big(\frac{\sqrt{fg}}{z^2}\,\partial_z\phi\Big)
= \Big(\frac{\omega^2}{z^2\sqrt{fg}} + \frac{k^2 q}{z^2} + \frac{m^2 q}{z^4}\Big)\phi,
\label{eq:radial}
\bbq
with $q=\sqrt{f/g}$. Near the boundary every solution behaves as
\bbe
\phi(z) \simeq c_-\, z^{\Delta_-} + c_+\, z^{\Delta_+},
\qquad
\Delta_\pm = \tfrac{3}{2} \pm \sqrt{\tfrac{9}{4} + m^2}.
\label{eq:falloff}
\bbq
The two point function of the bulk follows from the ratio of this two coefficients $\Delta_\pm$.
For $\Delta > 3/2$ the source is $c_-$ and the response is $c_+$. But for $\Delta < 3/2$ the roles are exchanged. This can be thought of as an alternate quantization branch. The Ising spin operator has $\Dsig = 0.5181489$ \cite{Kos:2016ysd} and lies on that branch. The energy operator has $\Deps = 1.412625$, which also satisfies $\Delta < 3/2$ belongs to the same branch. Thus both operators use alternate quantization. We keep this explicit as it exchanges source and response relative to the standard case. It also flips the sign of the response.

%% =====================================================================
\textit{Method.}---%
We parametrize the two metric functions by a small neural networks $u$,
\bbe
f(z) = e^{z^2 u_f(z)}, \qquad g(z) = e^{z^2 u_g(z)},
\label{eq:param}
\bbq
such that $f,g > 0$ and $f,g \to 1$ at the boundary.
Each $u$ is a two layer network with $32$ number of neurons. No other structure is imposed and 
the geometry is also not required to solve any equation of motion.

For a trial geometry,  we integrate \eqref{eq:radial} from an infrared wall to the boundary using a fourth order Runge Kutta scheme.
We can read off $c_\pm$ and form the model correlators. The loss function is the relative squared error against the lattice data.
Because the solver is differentiable, we can back propagate the loss through the integration and update the metric by gradient descent.
 We never impose the Einstein equations. The geometry is fixed entirely by the requirement of matching the boundary data. Gradients are propagated through the solver this replaces the Einstein equation by \emph{the boundary data} and act as the source of the geometry.

Lattice units and bulk units differ by an overall scale. Near criticality the data is close to scale free. Thus the scale is a gauge choice. We fix it by convention. The highest measured mode is mapped to a fixed bulk frequency. The normalization and contact term of each channel. They are fitted on the ultraviolet third ( top $\sim 33\%$ highest-momentum data ) against pure $\ads_4$. Details are in the Supplemental Material.

%% =====================================================================
\textit{Lattice data.}---%
We simulate the Ising model on an cubic lattice of size  $L^3$  at $\beta_c = 0.2216544$ \cite{Deng:2003wv, Ferrenberg:2018zst} using the Wolff cluster algorithm \cite{Wolff:1988uh}. We thermalize for a fixed number of sweeps. One sweep corresponds to L³ flipped spins. Measuring thermalization in sweeps rather than in cluster updates ensures that the thermalization length scales with the lattice size. Since we know individual Wolff clusters vary greatly in size near criticality. We measure two fields per configuration.
The spin field gives the $\sigma$ channel and the
mean subtracted bond energy field,
\bbe
\varepsilon(x) = \sigma(x)\sum_\mu \sigma(x+\hat{e}_\mu)
- \langle \sigma \sigma \rangle,
\label{eq:eps}
\bbq
gives the $\epsilon$ channel. Only the $k_z=0$ plane enters the fit. We therefore sum over the third spatial direction before transforming to momentum space.

As a check,  we extract the anomalous dimension from the measured $\sigma$ correlator. We find $\eta_{\rm fit} = 0.069$ at $L=128$ against the bootstrap value
$0.0363$. The corresponding fit for $\epsilon$ gives $1.826$ against $1.4126$. The offset in the $\epsilon$ channel is due to corrections to scaling rather than measurement error, as discussed below.

\begin{figure}[t]
\includegraphics[width=0.8\linewidth]{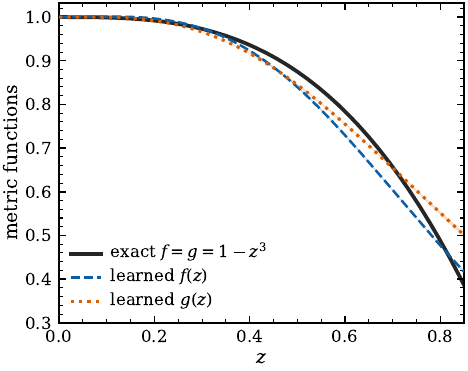}
\caption{Validation on synthetic data.
The machine is given correlators generated from
$f=g=1-z^3$ and must recover them.
Bands are the spread over three seeds.
The recovered cosmological constant is
$\Lambda = -3.009 \pm 0.001$ against the exact value $-3$.}
\label{fig:control}
\end{figure}

%% =====================================================================
\textit{Validation.}---%
We first validate the method on a known geometry.
We generate correlators from the $\ads_4$ Schwarzschild solution $f=g=1-z^3$ and hide the answer from the network. The reconstruction procedure recovers the metric to to better than one percent near the boundary and to a few percent in the interior. The cosmological constant is returned as $\Lambda = -3.009 \pm 0.001$, consistent with the exact value $\Lambda = -3$ (Fig.~\ref{fig:control}). This test also sets the resolution of the method. The reconstruction is sharp for $z \lesssim 0.3$ and degrades at larger depth.

\begin{table}[t]
\caption{Validation residuals for the two channels on a single $L=128$ dataset of $8000$
measurements. Diagonal entries are fits and the off-diagonal entries are predictions of the geometry trained on the other channel. All the values quoted here  are  the  medians over accepted seeds. }
\label{tab:main}
\begin{ruledtabular}
\begin{tabular}{lcc}
trained on & $\vsig$ & $\veps$ \\
\hline
$\sigma$ and $\epsilon$ & $8.9\times10^{-4}$ & $2.9\times10^{-4}$ \\
$\sigma$ only           & $1.9\times10^{-4}$ & $3.6\times10^{-3}$ \\
$\epsilon$ only         & $4.7\times10^{-2}$ & $9.2\times10^{-5}$ \\
\end{tabular}
\end{ruledtabular}
\end{table}

%% =====================================================================
\textit{Results.}---%
We now turn to the Ising data. We perform three fits on one dataset. The first uses both channels and one shared geometry. The second uses $\sigma$ alone and
the third uses $\epsilon$ alone. In the single channel fits the unused channel is never seen by the
optimizer. Its residual is a prediction.

Table~\ref{tab:main} collects the result.
Each channel fits better on its own than in the joint fit. The improvement is a factor of $4.7$ for $\sigma$ and $3.2$ for $\epsilon$. The joint geometry is a compromise and it costs both operators.

The off diagonal entries are larger. A geometry trained on $\epsilon$ miss predicts $\sigma$ by
more than two orders of magnitude. A geometry trained on $\sigma$ miss predicts $\epsilon$ by more than an order of magnitude. The two geometries are not small deformations of each other.

\begin{figure}[t]
\includegraphics[width=0.8\linewidth]{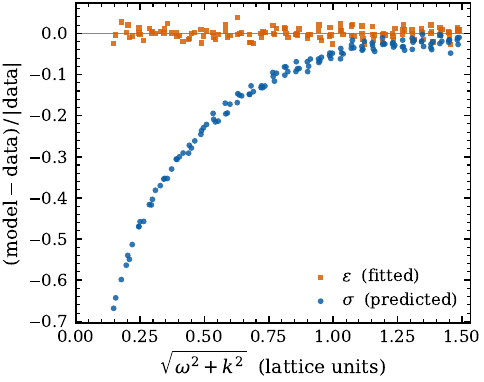}
\caption{Relative deviation per mode for the geometry trained on $\epsilon$ alone.
Squares are the fitted channel.
Circles are the predicted channel.
The fitted channel is flat. The predicted channel departs systematically at small momentum.}
\label{fig:residual}
\end{figure}

Figure~\ref{fig:residual} shows where the disagreement lives. The $\epsilon$ only channel fit the $\epsilon$ residuals are scattered about
zero at the level of two percent with no structure.
The predicted $\sigma$ residuals form a smooth arc.
They reach $-70$ percent at low momentum and approach zero at high momentum. The mismatch is not noise. It has a definite momentum dependence.

\begin{figure*}[t]
\includegraphics[width=0.8\textwidth]{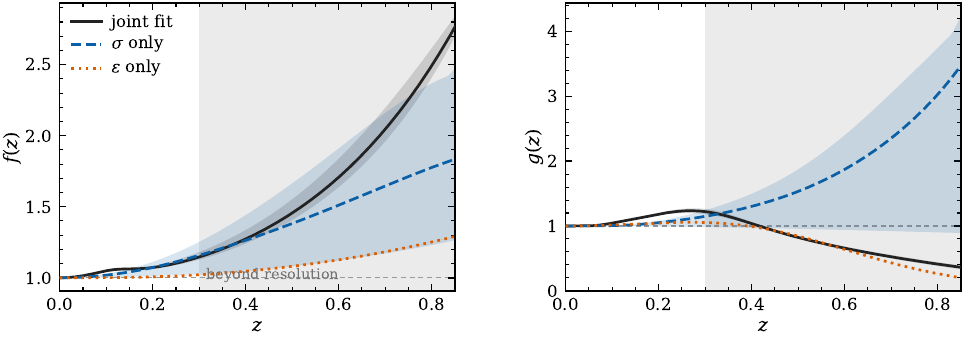}
\caption{The three reconstructed geometries.
Solid is the joint fit. Dashed is $\sigma$ only.
Dotted is $\epsilon$ only. Bands are seed to seed spread. The shaded region lies beyond the resolution set by Fig.~\ref{fig:control}.}
\label{fig:geoms}
\end{figure*}

%% =====================================================================
\textit{Robustness.}---%
Three checks control the obvious systematics.

First we vary the momentum window. The lowest modes carry the largest finite size contamination.
We repeat the analysis with one more mode removed.
Second we vary the lattice size.We repeat the analysis at $L=64$. A finite size artifact would shrink. Third we vary the optimizer. Every fit uses five seeds with warm restarts and identical settings.
Seeds whose fitted unit scale drifts by more than fifty percent from the ultraviolet anchor are discarded. The criterion is fixed in advance.
It removes one seed out of five in the $\sigma$ fit.

Table~\ref{tab:robust} collects the outcome.
The joint fit gives $\Lambda = -2.779 \pm 0.007$.
The mismatch persists in every variation.
At $L=64$ the factor is reduced, consistent with finite size rounding. This addresses the offset in the measured $\Deps$. The offset is a corrections to scaling effect of the energy operator.
So, it does not generate the tension.

\begin{table}[t]
\caption{Robustness of the channel mismatch.
Each row is a geometry trained on one channel. The fitted channel is in bold. The other residual is a prediction of that geometry. The last column is the best fit cosmological constant. Errors are the spread over accepted seeds.}
\label{tab:robust}
\begin{ruledtabular}
\begin{tabular}{llcccc}
$L$ & cut & trained on & $\vsig$ & $\veps$ & $\Lambda$ \\
\hline
128 & 2 & $\sigma$   & $\mathbf{1.9\times10^{-4}}$ & $3.6\times10^{-3}$ & $-3.24 \pm 0.42$ \\
128 & 2 & $\epsilon$ & $4.5\times10^{-2}$ & $\mathbf{9.5\times10^{-5}}$ & $-2.75 \pm 0.04$ \\
128 & 3 & $\sigma$   & $\mathbf{5.8\times10^{-4}}$ & $4.1\times10^{-3}$ & $-3.07 \pm 0.33$ \\
128 & 3 & $\epsilon$ & $2.4\times10^{-2}$ & $\mathbf{1.8\times10^{-4}}$ & $-2.81 \pm 0.03$ \\
64  & 2 & $\sigma$   & $\mathbf{8.4\times10^{-4}}$ & $3.0\times10^{-3}$ & $-3.93 \pm 1.54$ \\
64  & 2 & $\epsilon$ & $2.9\times10^{-2}$ & $\mathbf{8.2\times10^{-4}}$ & $-2.60 \pm 0.63$ \\
\end{tabular}
\end{ruledtabular}
\end{table}

%% =====================================================================
\textit{Discussion.}---%
In general relativity the metric is universal.
Every probe moves on the same geometry. Our measurement says that the two lightest Ising operators do not. Each prefers a different effective metric.

This is what one expects if the bulk carries more than a metric. A tower of higher spin fields would do it \cite{Klebanov:2002ja,Maldacena:2011jn}.
Different operators couple to different combinations of the tower. Each then sees a different effective geometry. Our data is consistent with that picture.
It does not identify the theory.

The result also has a negative reading which we state plainly. A single classical metric is not a good description of the Ising bulk at
the precision we reach. This is consistent with the criteria of \cite{Heemskerk:2009pn}. The central charge of the Ising model is of order unity
\cite{Chang:2024whx}. No semiclassical bulk was expected. What is new is a number attached to the failure.

We also attempted the spin two channel. The transverse traceless graviton obeys the massless scalar equation at zero spatial momentum and is dual to the stress tensor with $\Delta = 3$. We measured the bond anisotropy correlator with $10^4$ configurations. The signal is present. It is not precise enough to constrain a geometry. The residual is two orders of magnitude above the scalar channels.
We report this as a limitation and as a target.
Details are in the Supplemental Material.

Two future directions follow from here.
One is the finite momentum shear sector, which supplies more data per configuration.
The other is a backreacting bulk in which the metric and a scalar solve coupled equations.
Both would sharpen the statement made here.

%% =====================================================================
\begin{acknowledgments}
\textit{Acknowledgments}---%
The author thanks Prof. O. Parrikar  and Dr. Chintan Patel for their suggestive criticism on the eariler version of the draft. Author also wants to thank Pruthvi Suriya devara, Avijit Sinha and Yogeesh Reddy for their suggestions related to the ML part of this project. 
Computations were performed on Google Colab and on local hardware.
\end{acknowledgments}

\end{document}